\documentclass[conference,comsoc]{IEEEtran}
\IEEEoverridecommandlockouts

\usepackage{cite}
\usepackage{amsmath,amssymb,amsfonts}
\usepackage{algorithmic}
\usepackage{graphicx}
\usepackage{subfigure}
\usepackage{textcomp}
\usepackage{acronym}
\usepackage{multirow}
\newcommand{\minitab}[2][l]{\begin{tabular}{#1}#2\end{tabular}}

\usepackage{prettyref}
\newrefformat{fig}{Fig.~\ref{#1}}
\newrefformat{conj}{Conjecture~\ref{#1}}
\newrefformat{tab}{Table~\ref{#1}}
\newrefformat{sec}{Section~\ref{#1}}
\newrefformat{subsec}{Section~\ref{#1}}
\newrefformat{subsubsec}{Section~\ref{#1}}
\usepackage[super]{nth}
\usepackage[binary-units]{siunitx}
\usepackage{amsthm}

\def\BibTeX{{\rm B\kern-.05em{\sc i\kern-.025em b}\kern-.08em
    T\kern-.1667em\lower.7ex\hbox{E}\kern-.125emX}}
 
\theoremstyle{definition}

\theoremstyle{remark}

\acrodef{ofdm}[OFDM]{orthogonal frequency division multiplexing}
\acrodef{fbmc}[FBMC]{filter bank multicarrier}
\acrodef{fmt}[FMT]{filtered multitone}
\acrodef{ufmc}[UFMC]{universal filtered multicarrier}
\acrodef{gfdm}[GFDM]{generalized frequency division multiplexing}
\acrodef{zf}[ZF]{zero forcing}
\acrodef{mimo}[MIMO]{multiple-input-multiple-output}
\acrodef{siso}[SISO]{single-input single-output}
\acrodef{np}[NP]{nondeterministic polynomial}
\acrodef{sc}[SC]{single carrier}
\acrodef{sc-fde}[SC-FDE]{\ac{sc}-\ac{fde}}
\acrodef{evm}[EVM]{error vector magnitude}
\acrodef{dft}[DFT]{discrete Fourier transform}
\acrodef{ifft}[IFFT]{inverse fast Fourier transform}
\acrodef{cp}[CP]{cyclic prefix}
\acrodef{papr}[PAPR]{peak-to-average power ratio}
\acrodef{qam}[QAM]{quadrature amplitude modulation}
\acrodef{ask}[ASK]{amplitude shift keying}
\acrodef{psk}[PSK]{phase shift keying}
\acrodef{apsk}[APSK]{amplitude phase shift keying}
\acrodef{fsk}[FSK]{frequency shift keying}
\acrodef{gmsk}[GMSK]{gaussian minimum shift keying}
\acrodef{im}[IM]{index modulation}
\acrodef{fft}[FFT]{fast Fourier transform}
\acrodef{ici}[ICI]{inter-carrier interference}
\acrodef{isi}[ISI]{inter-symbol interference}
\acrodef{ini}[INI]{inter-numerology interference}
\acrodef{pdf}[PDF]{probability density function}
\acrodef{cr}[CR]{cognitive radio}
\acrodef{wola}[WOLA]{weighted overlap-add}
\acrodef{qos}[QoS]{Quality of Service}
\acrodef{csi}[CSI]{channel state information}
\acrodef{ber}[BER]{bit-error rate}
\acrodef{rf}[RF]{radio frequency}
\acrodef{rrc}[RRC]{root-raised cosine}
\acrodef{as}[AS]{alignment signal}
\acrodef{snr}[SNR]{signal-to-noise ratio}
\acrodef{dfe}[DFE]{decision-feedback equalizer}
\acrodef{fde}[FDE]{frequency domain equalization}
\acrodef{oob}[OOB]{out-of-band}
\acrodef{aci}[ACI]{adjacent channel interference}
\acrodef{phy}[PHY]{physical layer}
\acrodef{mse}[MSE]{mean-square error}
\acrodef{pa}[PA]{power amplifier}
\acrodef{rb}[RB]{Resource Block}
\acrodef{ran}[RAN]{radio-access network}
\acrodef{rat}[RAT]{radio access technology}
\acrodefplural{rat}[RATs]{radio access technologies}
\acrodef{mmwave}[mmWave]{millimeter wave}
\acrodef{3gpp}[3GPP]{Third Generation Partnership Project}
\acrodef{sc}[SC]{single carrier}
\acrodef{uw}[UW]{Unique Word}
\acrodef{zt}[ZT]{zero tail}
\acrodef{gi}[GI]{guard interval}
\acrodef{sinr}[SINR]{signal to interference plus noise ratio}
\acrodef{dpss}[DPSS]{discrete prolate spheroidal sequences}
\acrodef{bler}[BLER]{block error rate}
\acrodef{fdd}[FDD]{frequency domain duplexing}
\acrodef{tdd}[TDD]{time domain duplexing}
\acrodef{ul}[UL]{uplink}
\acrodef{dl}[DL]{downlink}
\acrodef{bs}[BS]{base station}
\acrodef{af}[AF]{ambiguity function}
\acrodef{ms}[MS]{mobile station}
\acrodef{iot}[IoT]{Internet of Things}
\acrodef{pdp}[PDP]{power delay profile}
\acrodef{ftn}[FTN]{Faster-than-Nyquist}
\acrodef{svd}[SVD]{singular-value decomposition}
\acrodef{uw-ofdm}[UW-OFDM]{\ac{uw}-\ac{ofdm}}
\acrodef{dft-s-ofdm}[DFT-s-OFDM]{\ac{dft}-spread-\ac{ofdm}}
\acrodef{lte}[LTE]{Long Term Evolution}
\acrodef{lte-a}[LTE-A]{\ac{lte}-Advanced}
\acrodef{gsm}[GSM]{Global System for Mobile Communications}
\acrodef{ma}[MA]{multiple accessing}
\acrodef{oma}[OMA]{orthogonal multiple accessing}
\acrodef{noma}[NOMA]{non-orthogonal multiple accessing}
\acrodef{cdnoma}[CD-NOMA]{Code domain \ac{noma}}
\acrodef{pdnoma}[PD-NOMA]{Power domain \ac{noma}}
\acrodef{fdma}[FDMA]{frequency division multiple accessing}
\acrodef{ofdma}[OFDMA]{orthogonal frequency division multiple accessing}
\acrodef{sc-fdma}[SC-FDMA]{single carrier-frequency division multiple accessing}
\acrodef{cdma}[CDMA]{code division multiple accessing}
\acrodef{tdma}[TDMA]{time division multiple accessing}
\acrodef{embb}[eMBB]{enhanced-mobile broadband}
\acrodef{mmtc}[mMTC]{massive machine type communications}
\acrodef{urllc}[uRLLC]{ultra reliable and low latency communications}
\acrodef{dvb}[DVB]{digital video broadcasting}
\acrodef{comp}[CoMP]{Coordinated Multi Point}
\acrodef{jt-comp}[JT-CoMP]{Joint Transmission-\ac{comp}}
\acrodef{sic}[SIC]{successive interference cancellation}
\acrodef{ml}[ML]{maximum likelihood}
\acrodef{sl}[SL]{sidelink}
\acrodef{qpsk}[QPSK]{quadrature phase shift keying}
\acrodef{mpa}[MPA]{message passing algorithm}
\acrodef{amc}[AMC]{adaptive modulation and coding}
\acrodef{umts}[UMTS]{Universal Mobile Telecommunications Service}
\acrodef{e-utra}[E-UTRA]{Evolved \ac{umts} Terrestrial Radio Access}
\acrodef{ue}[UE]{user equipment}
\acrodef{harq}[HARQ]{Hybrid Automatic Repeat Request}
\acrodef{nr}[NR]{new radio}
\acrodef{hetnet}[HetNet]{Heterogeneous network}
\acrodef{pot}[POT]{partially overlapping tones}
\acrodef{rms}[RMS]{root mean square}
\acrodef{must}[MUST]{multiuser superposition transmission}
\acrodef{csma}[CSMA]{carrier sense \ac{ma}}
\acrodef{sdma}[SDMA]{space division \ac{ma}}
\acrodef{mno}[MNO]{mobile network operator}
\acrodef{d2d}[D2D]{device-to-device}
\acrodef{v2v}[V2V]{vehicle-to-vehicle}
\acrodef{v2x}[V2X]{vehicle-to-infrastructure}
\acrodef{med}[MED]{maximum excess delay}
\acrodef{awgn}[AWGN]{additive white Gaussian noise}
\acrodef{cfr}[CFR]{channel frequency response}
\acrodef{cir}[CIR]{channel impulse response}
\acrodef{sir}[SIR]{signal-to-interference ratio}
\acrodef{lut}[LUT]{lookup table}
\acrodef{ssw}[SSW]{subcarrier specific window}
\acrodef{cqi}[CQI]{channel quality indicator}
\acrodef{bpsk}[BPSK]{binary phase shift keying}
\acrodef{mcs}[MCS]{modulation and coding scheme}
\acrodef{dac}[DAC]{digital-to-analog converter}
\acrodef{adc}[ADC]{analog-to-digital converter}
\acrodef{e2e}[E2E]{end-to-end}
\acrodef{kpi}[KPI]{key performance indicator}
\acrodef{p1db}[P1dB]{1 dB compression point}
\acrodef{sqnr}[SQNR]{signal-to-quantization-noise ratio}
\acrodef{cmos}[CMOS]{complementary metal–oxide–semiconductor}
\acrodef{cs}[CS]{cyclic suffix}
\acrodef{pn}[PN]{phase noise}
\acrodef{xo}[XO]{crystal oscillator}
\acrodef{dc}[DC]{direct current}
\acrodef{ac}[AC]{alternating current}
\acrodef{ss}[SS]{spread spectrum}
\acrodef{dsss}[DSSS]{direct sequence \ac{ss}}
\acrodef{fhss}[FHSS]{frequency hopping \ac{ss}}
\acrodef{gps}[GPS]{global positioning system}
\acrodef{ap}[AP]{access point}
\acrodef{rts}[RTS]{request to send}
\acrodef{cts}[CTS]{clear to send}
\acrodef{macaw}[MACAW]{multiple access collision avoidance for wireless}
\acrodef{sifs}[SIFS]{short interframe space}
\acrodef{ssid}[SSID]{service set identifier}
\acrodef{mu}[MU]{multi-user}
\acrodef{sta}[STA]{station}
\acrodef{wcsp}[WCSP]{Wireless Communications and Signal Processing}
\acrodef{metu}[METU]{Middle East Technical University}
\acrodef{njit}[NJIT]{New Jersey Institute of Technology}
\acrodef{nnrp}[NNRP]{network normalized received power}
\acrodef{mrc}[MRC]{maximum ratio combining}
\acrodef{uci}[UCI]{uplink channel information}
\acrodef{map}[MAP]{maximum a posteriori}
\acrodef{ls}[LS]{least squares}
\acrodef{mai}[MAI]{multiple access interference}
\acrodef{bw}[BW]{bandwidth}
\acrodef{bwp}[BWP]{\ac{bw} Part}
\acrodef{ba}[BA]{\ac{bw} adaptation}
\acrodef{sa}[SA]{standalone}
\acrodef{nsa}[NSA]{non-\ac{sa}}
\acrodef{ffs}[FFS]{for further study}
\acrodef{scs}[$\Delta{f}$]{subcarrier spacing}
\acrodef{cpdur}[$T_\mathrm{CP}$]{\ac{cp} duration}
\acrodef{reg}[REG]{Resource Element Group}
\acrodef{tti}[TTI]{Transmission Time Interval}
\acrodef{fr}[FR]{frequency range}
\acrodef{mac}[MAC]{medium access control}
\acrodef{gnb}[gNB]{next generation node B}

\def\BibTeX{{\rm B\kern-.05em{\sc i\kern-.025em b}\kern-.08em
    T\kern-.1667em\lower.7ex\hbox{E}\kern-.125emX}}
\begin{document}

\title{Fundamentals of Multi-Numerology 5G New Radio
\thanks{The extended and updated version of this work was published in River Publishers Journal of Mobile Multimedia \cite{yazar2018b}. https://doi.org/10.13052/jmm1550-4646.1442}}

\author{
\IEEEauthorblockN{Ahmet Yazar\IEEEauthorrefmark{1}, Berker Pek\"{o}z\IEEEauthorrefmark{2} and H\"{u}seyin Arslan\IEEEauthorrefmark{1}\IEEEauthorrefmark{2}}
\IEEEauthorblockA{\IEEEauthorrefmark{1}Department of Electrical and Electronics Engineering, Istanbul Medipol University, Istanbul, 34810 Turkey\\
Email: \{ayazar,huseyinarslan\}@medipol.edu.tr}
\IEEEauthorblockA{\IEEEauthorrefmark{2}Department of Electrical Engineering, University of South Florida, Tampa, FL 33620 USA\\
Email: \{pekoz,arslan\}@usf.edu}
}

\maketitle

\begin{abstract}
The physical layer of 5G cellular communications systems is designed to achieve better flexibility in an effort to support diverse services and user requirements. OFDM waveform parameters are enriched with flexible multi-numerology structures. This paper describes the differences between \ac{lte} systems and \ac{nr} from the flexibility perspective. Research opportunities for multi-numerology systems are presented in a structured manner. Finally, \ac{ini} results as a function of guard allocation and multi-numerology parameters are obtained through simulation.
\end{abstract}

\acresetall



\section{Introduction}
\label{sec:introduction}

\ac{lte} waveform has a fixed structure that is optimized to serve high data rate applications. There is only limited support for other applications due to the inflexibility of the waveform. An example for the limited flexibility is the extended \ac{cp} configuration utilized by macrocell \acp{bs} at all times to keep the system operating at larger delay spreads at the cost of reduced spectral efficiency \cite{3gpp.36.104}. This adaptation is rather limited as the configuration is static; even when not needed by any \ac{ue}, the system is configured to operate with these parameters and does not have the flexibility to improve the efficiency by utilizing normal \ac{cp}. Other than delay spread, any degradation in \ac{sinr}, regardless of the cause, is addressed solely using \ac{amc} by reducing the throughput until a fixed reliability threshold is achieved \cite{3gpp.36.211}. For instance, if \ac{sinr} degrades due to \ac{ici} in high mobilities, this issue can only be addressed using \ac{amc} in \ac{lte}, reducing throughput and under-utilizing the \ac{bw}. As can be seen from the above examples, \ac{lte} has limited flexibility and cannot support the rich application and user requirements of 5G services \cite{yazar2018aflexibility}.

5G is designed to provide a wide variety of services by rendering waveform parameters flexibly\cite{ankarali2017flexible}. The new design paradigms make an \ac{embb} experience possible everywhere, including highly mobile \ac{ue} connected to macrocells. The flexibilities introduced to the waveform enable reduced latencies and improved reliability, empowering \ac{urllc} rather than high data rate applications. In addition, \ac{mmtc} is enabled for suitable scenarios with \ac{nr}.

This flexibility was provided by coexisting of numerologies, where each numerology consists of a set of parameters defining the frame and lattice structure of the waveform. In contrast to the single-numerology utilization in \ac{lte}, \ac{nr} allows simultaneous multi-numerology utilization \cite{3gpp.38.211}. One of the first studies that incorporated multi-numerology or mixed-numerology systems and designed a framework that provides numerous services simultaneously in a unified frame was \cite{sahin2012multiuser}. Multi-numerology structures that were included in the \ac{3gpp} \ac{nr} standardization were also studied in literature \cite{sec1:yoshihisa1, sec1:yoshihisa2, sec1:ericsson, sec1:tafazolli1, sec1:tafazolli2}.

In this paper, three main contributions have been made as listed below:
\begin{enumerate}
  \item \ac{lte} and \ac{nr} were compared from the flexibility perspective regard to \ac{3gpp} 38-series documents.
  \item Research opportunities for multi-numerology systems are presented in a structured manner.
  \item Through simulation, \ac{ini} results as a function of guard allocation and multi-numerology parameters are obtained.
\end{enumerate}

The rest of the paper is organized as follows: \prettyref{sec:flexibility} presents the comparison between \ac{lte} and \ac{nr} systems from the flexibility perspective. New concepts introduced in \ac{nr} are also described in this section regarding to \ac{3gpp} 38-series documents. Research opportunities for potential improvements of multi-numerology systems are explained in \prettyref{sec:opportunity}. \prettyref{sec:simulation} shows simulation results for multi-numerology structures. Finally, the conclusion is given in \prettyref{sec:conclusion}.


\section{Flexibility of \ac{nr} compared to \ac{lte}}
\label{sec:flexibility}

New concepts are introduced and building blocks are defined to provide more flexible \acp{rat} in \cite{intel2018} and \cite{ericsson2017}. In this section, concepts that were introduced in \ac{nr} are defined and their differences with \ac{lte} are distinguished. Release 15 was taken as the reference for both \ac{nr} and \ac{lte}.

In \ac{3gpp} Release 15, \ac{sa} operation according to \cite{3gpp.38.101-1, 3gpp.38.101-2} and \ac{nsa} operation coexisting with other technologies according to \cite{3gpp.38.101-3} are defined. \ac{nsa} operation was finalized in Release 15, but some issues regarding \ac{sa} operation, along with details necessary to provide \ac{mmtc}, was left for further study to be finalized in Release 16.

Waveform defines how the resources are placed in the time-frequency lattice and the structure (pulse shapes and filters) that maps information symbols to these resources \cite{sahin2012multiuser}. In the \ac{dl}, \ac{nr} uses \ac{cp}-\ac{ofdm} with multi numerologies (a mother waveform plus its derivatives). The mother waveform is the same in \ac{lte} but there is only one numerology. In the \ac{ul}, there is an option to use either of \ac{cp}-\ac{ofdm} and \ac{dft-s-ofdm} with multi numerologies for \ac{nr} \cite{3gpp.38.804}. However, the only option in \ac{lte} is \ac{dft-s-ofdm} with a single numerology.

The time-frequency lattice is the grid of discrete resources in the continuous time-frequency plane, where each ``atom'' on the grid shows where the continuous plane has been sampled, thus defining where/when information can be transmitted\cite{sahin2012multiuser}. \ac{lte} used a fixed lattice in which the frequency (and corresponding time) spacing between each point was always the same throughout the whole transmission band \cite{ankarali2017flexible}. However, \ac{nr} defines flexible time-frequency lattice enabling multi-numerology structure. For the case of \ac{ofdm}, numerology set consists of number of subcarriers, \ac{scs}, slot duration and \ac{cpdur} \cite{yazar2018aflexibility}. The \ac{scs}, \ac{cpdur}, slot duration, and maximum \ac{bw} allocation options for \ac{nr} numerologies according to \cite{3gpp.38.211} and \cite{3gpp.38.104} are presented in \prettyref{tab:numerology}. These numerologies can be used simultaneously in a cell. On the contrary, \ac{lte} is a single-numerology system thus all these parameters are fixed at all times for a \ac{bs}.

A \ac{bwp} is a new term that defines a fixed band over which the communication taking place uses the same numerology throughout the existence of the \ac{bwp} \cite{3gpp.38.213}. It is a bridge between the numerology and scheduling mechanisms. \acp{bwp} are controlled at the \ac{bs} based on \ac{ue} needs and network requirements. In contrast to \ac{lte}, 5G \acp{ue} need not monitor the whole transmission \ac{bw}; they only scan the \acp{bwp} assigned to themselves. \acp{bwp} allow \acp{ue} to process only part of the band that contain their symbols, reducing power consumption and enabling longer battery lives. This is very useful for the low-power communications systems, particularly \ac{mmtc} services. \acp{bwp} may overlap to facilitate low latency services while providing data to noncritical services to ensure efficient utilization of resources.

\ac{bs} channel \ac{bw} is another new term that refers to the contiguous \ac{bw} currently in use by the \ac{gnb} for either transmission or reception \cite{3gpp.38.817-01}. In other words, it refers to the total \ac{bw} that is processed by the \ac{gnb}.

Unlike \ac{lte} slots that consist of 7 \ac{ofdm} symbols in case of normal \ac{cp}, \ac{nr} slots can consist of 14 symbols for \acp{scs} up to \SI{60}{\kilo\hertz} \cite{3gpp.38.802}. Furthermore, \ac{lte} \acp{rb} cover 12 consecutive subcarriers over a subframe (i.e., two slots) duration, whereas \ac{nr} \acp{rb} are defined only using the same \ac{bw} definition; their durations are not fixed \cite{3gpp.38.211}. As opposed to the fixed \ac{lte} \ac{tti} duration of one slot, \ac{nr} \ac{tti} may be a mini-slot in the case of \ac{urllc} or beam-sweeping operation in frequency range-2, a slot for regular operation, or multiple slots in the case of large number of \ac{embb} packets; thus having a definition varying as a function of the service \cite{3gpp.38.802}. \ac{nr} re-uses the \ac{lte} radio frame definition \cite{3gpp.38.201}, however, the number of slots per sub-frame depends on the \ac{scs} and is given by the multiplicative inverse of the slot duration seen in \prettyref{tab:numerology} \cite{3gpp.38.211}.

\begin{table}
\renewcommand{\arraystretch}{1.3}
\caption{Numerology Structures and the Corresponding Maximum \ac{bw} Allocations for Data Channels in 5G}
\label{tab:numerology}
\centering
\begin{tabular}{|c|c|c|c|c|}
\hline
Frequency & \ac{scs} & \ac{cpdur} & Slot & Max. \ac{bw}\\
Range (FR) & $(\si{\kilo\hertz})$ & ($\si{\micro\second}$) & Duration (\si{\milli\second}) & (\si{\mega\hertz})\\
\hline
\multirow{ 3}*{\minitab[c]{FR-1}} & 15 & 4.76 & 1 & 50\\
\cline{2-5}
& 30 & 2.38 & 0.5 & 100\\
\cline{2-5}
& 60 & 1.19 $|$ 4.17 & 0.25 & 100\\
\hline
\multirow{ 2}*{\minitab[c]{FR-2}} & 60 & 1.19 $|$ 4.17 & 0.25 & 200\\
\cline{2-5}
 & 120 & 0.60 & 0.125 & 400\\
\hline
\end{tabular}
\end{table}


\section{Research Opportunities for Multi-Numerology Systems}
\label{sec:opportunity}

As it can be seen from the previous section, the main flexibility causative for \ac{nr} is mostly focused on the new frame with multi-numerology structures. Different user and service requirements can be met using multiple numerologies. In other words, multiplexing numerologies provides the flexibility needed by \ac{nr}.

This section presents exemplary multi-numerology algorithms that exploit the flexibilities in \ac{nr} design pointed out in \prettyref{sec:flexibility}. \ac{3gpp} standards give the \ac{bs} and \ac{ue} manufacturers the freedom to implement any additional algorithm they desire as long as it is transparent to the receiver\cite{3gpp.38.912}. Examples provided in this section also exploit this degree of freedom.

\subsection{Non-Orthogonality of Multi Numerologies}

\begin{figure}[t]
  \centering
  \includegraphics[width=8.5cm]{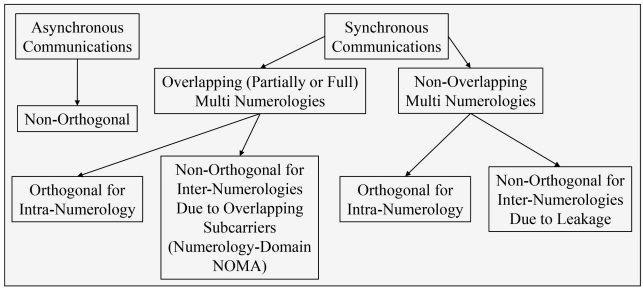}
  \caption{Orthogonality and non-orthogonality for intra-numerology and inter-numerologies cases.}
\label{fig:Fig2}
\end{figure}

Resource elements within the same numerology are orthogonal to each other, but resource elements of any two different numerologies are non-orthogonal to each other and interfere with one another \cite{ankarali2017flexible}. Non-orthogonality can result either from partially or completely overlapped numerologies, or non-overlapping numerologies for synchronous communications. As it can be seen from \prettyref{fig:Fig2}, non-orthogonality is originated from overlapping subcarriers for the first case. However, the reason of non-orthogonality is \ac{oob} emission in the second case. Optionally, guard bands can be employed to reduce interference for the second case. Performance analysis for the effects of guard bands between numerologies is given in \prettyref{sec:simulation}. Besides these, subcarriers are non-orthogonal to each other for intra- or inter-numerology domains in asynchronous communications \cite{seda2018}.

In \cite{ayman2018}, authors proposed a numerology-domain \ac{noma} system with overlapping multi-numerology structures. \ac{nr} allows overlapping of \acp{bwp} using different numerologies in time-frequency grid\cite{intel2018}. Numerology-domain \ac{noma} system designs can be developed to exploit this gap in 5G.

\subsection{Numerology Selection Methodologies}

\ac{bwp} is a useful tool for multi-numerology systems as \ac{bwp} defines a specific numerology. \ac{bs} can modify \ac{ue} numerologies by changing it's \acp{bwp}. Parameter configuration process for the \acp{bwp} is employed by \ac{ba} tool on \ac{bs} \cite{3gpp.38.300}. There can be up to four defined \acp{bwp} for each \ac{ue} but there is one active \ac{bwp} for each user in Release 15. However, future \ac{nr} releases are planned to allow multiple (up to four per \ac{ue} in Release 16) active \ac{bwp} configurations.

Active \acp{bwp} and the corresponding numerologies can be selected using different methodologies. Various trade-offs between distinct performance metrics that include spectral efficiency, \ac{ini}, flexibility, and complexity can be considered while deciding on active \acp{bwp} and so numerologies.

\begin{figure}[t]
  \centering
  \includegraphics[width=8.4cm]{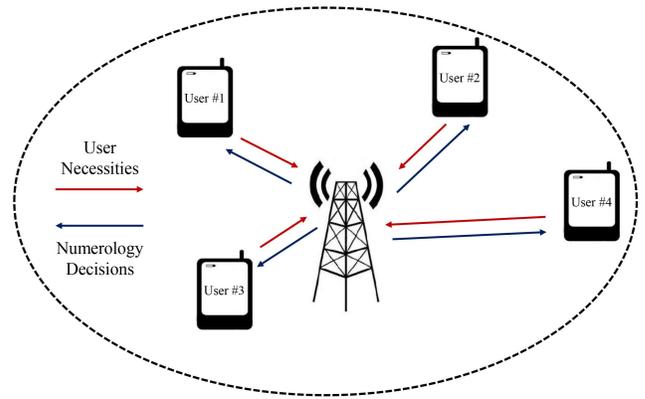}
  \caption{Simple representation of numerology selection methodology in a cell serving users with various necessities \cite{yazar2018aflexibility}. User necessities are gathered by \ac{bs} at different times but numerology decisions are made at the same time.}
\label{fig:Fig4}
\end{figure}

For one active \ac{bwp} at a time case, an example numerology selection methodology is proposed in \cite{yazar2018aflexibility} that uses a heuristic algorithm to configure numerologies suitable for each user. \prettyref{fig:Fig4} illustrates this resource allocation optimization methodology. The proposed method also provides an active \ac{bwp} switching mechanism. \ac{scs}, \ac{cpdur}, and spectral efficiency requirements of all users in a cell are input to the algorithm.

It is possible to increase the number of numerologies in beyond 5G. Offering more numerologies simultaneously ensures that all user and service requirements are satisfied, but this requires more sophisticated numerology selection mechanisms. To reduce computational costs, \acp{bs} may use two-step numerology selection methods in the future. The first step decides on the most suitable numerology set between different sets. Then, the second step determines the best numerologies from the set that is selected in the first step. Additionally, there can be many different numerology selection methods for multiple \acp{bwp} active at a time case.

\subsection{\ac{ini} Estimation Models}

\ac{ini} can be simply defined as \ac{ici} between subcarriers of different numerology structures. The amount of \ac{ini} can vary with \ac{scs}, \ac{bw}, guards, \ac{cpdur}, the number of different numerologies, alignment of different numerologies in frequency domain, filtering/windowing usage, frequency bands, user powers, and so on. All of these parameters need to be analyzed together to form estimation models for \ac{ini}. 

\ac{ini} estimation is very important topic because it can be used as a feedback to all other adaptive systems that include adaptive guards, numerology selection, filtering/windowing decision, and optimization of the number of numerologies. Interference models can be very useful for adaptive decision on different algorithms for multi-numerology systems. For example, an \ac{ini} estimation method between different transmitter and receiver windowed \ac{ofdm} numerologies are provided in \cite{pekoz2017adaptive}, where the exact calculation of \ac{ini} using the \acp{cir} and data of all users, as well as estimation techniques for practical cases such as unknown data as well as unknown \acp{cir} are provided.

\begin{figure*}[t]
  \centering
  \includegraphics[width=15.7cm]{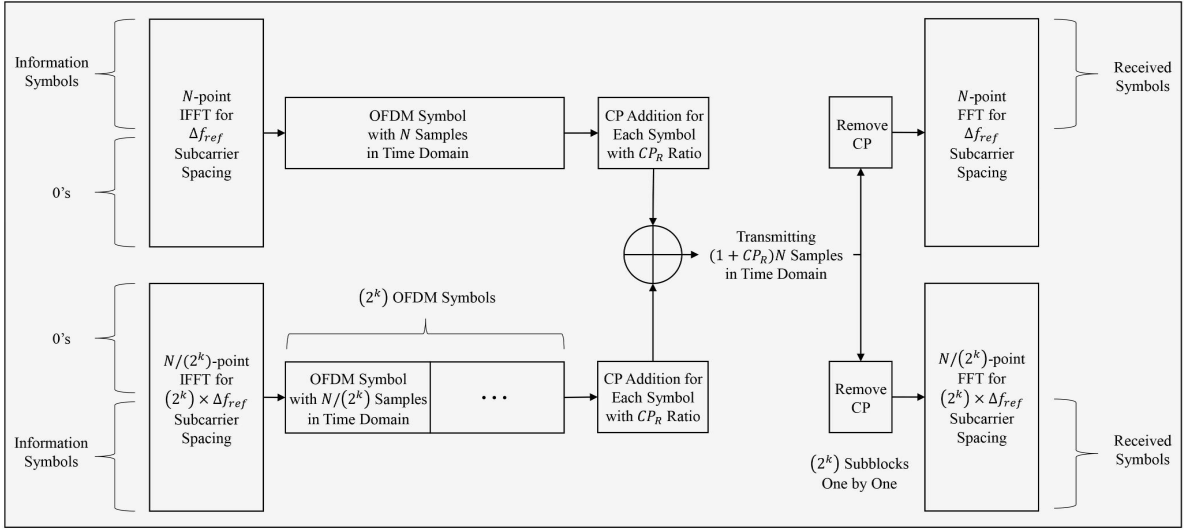}
  \caption{Block diagram for the simple implementation of multi numerologies. The scaling factor of \acp{scs} is chosen as $2^k$, where $k$ is a positive integer.}
\label{fig:Fig3}
\end{figure*}

\subsection{Effects of Guard Bands Between Multi Numerologies}

This subsection deals with the adjustment of frequency domain guards with respect to estimated \ac{ini} after numerologies are selected. In \ac{3gpp} standards, it is revealed that there are minimum guard band requirements, a maximum or an optimum value is not enforced, making guard bands choices flexible with high granularity \cite{3gpp.38.817-01}. Adaptive guard band concept for different numerologies becomes a crucial research area at this point.

As it is well known, the \ac{ofdm} signal is well localized in the time domain with a rectangular pulse shape, which corresponds to a sinc pulse in the frequency domain. Sincs cause significant \ac{oob} emission and guard bands are needed between two adjacent subbands with different numerologies to handle the interference.

The \ac{oob} emission increases as the symbol duration decreases. Therefore, more guard band is required for the numerologies with higher \ac{scs}. For the edge subcarriers of two adjacent numerologies, \ac{sinr} decrease is more significant compared to the remaining subcarriers. Most of the interference comes from the edge subcarriers \cite{Sahin2011}. Grouping services in \acp{bwp} reduces the amount of necessary guards and eases scheduling when such fast numerology variations become necessities. Moreover, passing \ac{ofdm} signal through power amplifiers causes non-linear distortions. The \ac{papr} and \ac{oob} emission increase as the number of active subcarriers increases. As a result, more guard band is needed for the transmissions with wider occupied numerology \acp{bw}.

In \prettyref{sec:simulation}, the effects of guard bands between multi numerologies with the performance analysis results are shown regarding to the implementation block diagram given in \prettyref{fig:Fig3}.

\subsection{Effects of Guard Intervals for \ac{ini} Elimination}

In addition to guard bands between different numerologies, the guards in time and frequency domains must be jointly optimized to boost the spectral efficiency \cite{demir2017theimpact}. Various slot configurations and \ac{ue} scheduling guidelines reveal that few restrictions exist regarding scheduling users in time domain. This implies that the guard times can also be utilized flexibly, similar to guard bands. Combining time-frequency guard flexibility yields that empty resource elements can virtually be placed anywhere. Interpreting this at a multi-user level reveals that the \ac{ul} slot of one \ac{ue} and the \ac{dl} slot of another \ac{ue} can be scheduled to consecutive time or frequency resources with little guard time and band. This poses serious requirements in pulse shaping, making localized pulses and interference rejection techniques critical.

Also, the use case and power imbalance factors should be considered on the guard allocation. The power control mechanism mitigates the interference problem in power imbalance scenarios as well, but it prevents deployment of higher order modulation for the users that experience higher \ac{sinr}. Thus, power control requires relaxation using an adaptive guard design to increase the throughput. The potential of adaptive guards can be increased further by utilizing an interference-based scheduling algorithm \cite{demir2017theimpact}.

\subsection{Filtering and Windowing in \ac{nr}}

\ac{ini} cannot only be handled using guards but also with the filtering and windowing approaches that require additional guards. Applying filters and windows methods are left for the implementation in \ac{3gpp} standardization.

Allocating users optimal guards minimizes but not completely eliminates the interference on the received signal in a non-orthogonal system. The receiver may also engage in filtering and windowing to further eliminate the remaining interference, but doing so using conventional methods requires additional guards. The assigned optimum guards may not even be sufficient if extreme latencies are required. The algorithm presented in \cite{pekoz2017adaptive} deals with the minimization of aggregate \ac{isi}, \ac{ici} and \ac{aci} by windowing each received subcarrier with the window function that minimizes the aggregate interference at that subcarrier. The optimal window lengths require perfect knowledge of the interfering users' data and channels. While this can be known and applied at \ac{ul} reception at the \ac{gnb} in a manner similar to \ac{sic} or \ac{mu} detection, this cannot be done at the \ac{ue}. Therefore, the algorithm presents methods to estimate optimal subcarrier specific window durations if only the \acp{cir}, \acp{pdp} or the power offsets of the interferers are known.

\subsection{Optimization on the Number of Active Numerologies}

Authors of \cite{yazar2018aflexibility} find the efficient number of active numerologies that should be simultaneously employed by users. The algorithm aims to minimize various overheads to provide a practical solution satisfying different service and user requirements using multi-numerology structures. All of the different numerologies that are defined in standards do not need to be used in every situation.

Basically, the amount of total guard band in the lattice increases with increasing number of numerologies. Hence, there is a trade-off between the spectral efficiency and multi-numerology system flexibility. Although not imposed by the standard\cite{3gpp.38.211}, they allocate \acp{bwp} configured to use the same numerologies consecutively in an effort to reduce guard bands and computational complexity.


\section{Simulation Results on Multi-Numerology}
\label{sec:simulation}

In this section, \ac{ini} results as a function of guard allocation and multi-numerology parameters are provided based on the block diagram in \prettyref{fig:Fig3}. It is assumed that \acp{bwp} with different numerologies are not overlapped at a time and \acp{bwp} with the same numerologies are grouped together in the frequency domain. Also, user powers are taken as equal.

Random \ac{bpsk} symbols are generated separately for two-numerology structure. For the first numerology, which has $\SI[parse-numbers = false]{\Delta f_\text{ref}}{\kilo\hertz}$
subcarrier spacing, $N$-point \ac{ifft} is employed.
The second numerology has $\SI[parse-numbers = false]{2^k\times\Delta{f}_{\text{ref}}}{\kilo\hertz}$ subcarrier spacing and uses $N/(2^k)$-point \ac{ifft}, where $2^k$ is the scaling factor and $k$ is a positive integer. Here, the second half of the \ac{ifft} inputs for the first numerology and the first half of the \ac{ifft} inputs for the second numerology are zero-padded to separate two numerologies in frequency domain. After each \ac{ifft} operation, \ac{cp} samples are added with a ratio of $CP_R$ to every \ac{ofdm} symbol. There are $2^k$ OFDM symbols with the second numerology corresponding to one OFDM symbol with the first numerology. Thus, the number of samples for each of the numerologies are the same, and they can be added to form a composite signal at the transmitter.

Wireless channel and noise are ignored to just focus on the \ac{ini} in the simulation results. At the receiver side, \ac{cp} samples are removed from each \ac{ofdm} symbol. $N$-point \ac{fft} is used for the first numerology over full composite signal. However, only $N/(2^k)$ samples of the composite signal to make them input into $N/(2^k)$-point \ac{fft} for the second numerology. $2^k$ subblocks are constituted by dividing the composite signal into $2^k$ parts and these subblocks are processed one by one. The first half of the \ac{fft} output for the first numerology and the second half of the \ac{fft} output for the second numerology are taken to obtain received symbols.

\begin{figure*}[t]
  \centering
  \subfigure[Case 1: Numerology-1 has \SI{15}{\kilo\hertz} \ac{scs} and Numerology-2 has \SI{30}{\kilo\hertz} \ac{scs}. Guard bands are \SIlist{0;180;360}{\kilo\hertz}.]{\includegraphics[width=8.1cm]{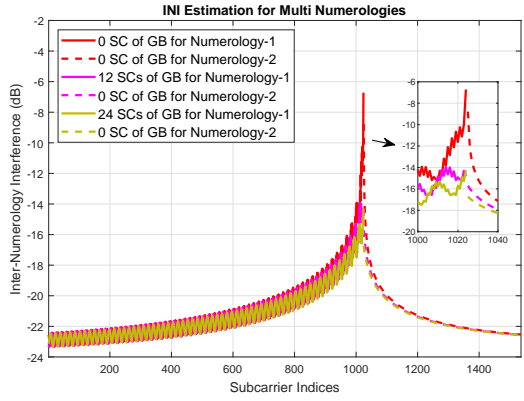}}\qquad
  \subfigure[Case 2: Numerology-1 has \SI{15}{\kilo\hertz} \ac{scs} and Numerology-2 has \SI{30}{\kilo\hertz} \ac{scs}. Guard bands are \SIlist{15;195;375}{\kilo\hertz}.]{\includegraphics[width=8.1cm]{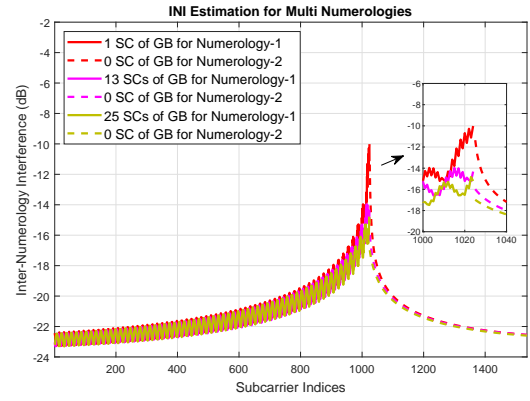}}\\
  \subfigure[Case 3: Numerology-1 has \SI{15}{\kilo\hertz} \ac{scs} and Numerology-2 has \SI{60}{\kilo\hertz} \ac{scs}. Guard bands are \SIlist{0;180;360}{\kilo\hertz}.]{\includegraphics[width=8.1cm]{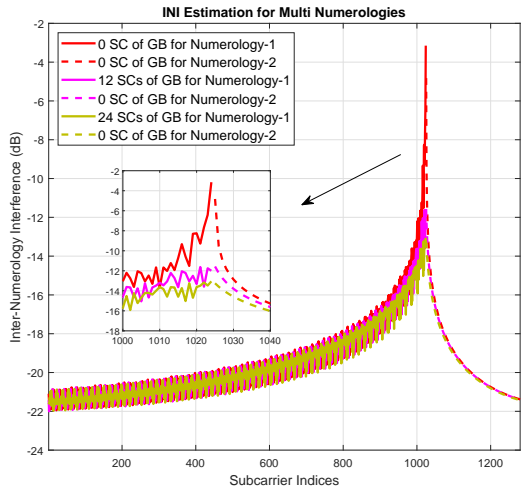}}\qquad
  \subfigure[Case 4: Numerology-1 has \SI{15}{\kilo\hertz} \ac{scs} and Numerology-2 has \SI{60}{\kilo\hertz} \ac{scs}. Guard bands are \SIlist{45;225;405}{\kilo\hertz}.]{\includegraphics[width=8.1cm]{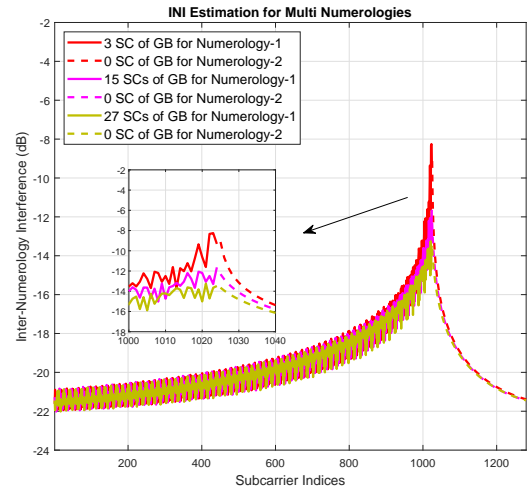}}
  \caption{Simulation results for four different cases with different guard band amounts between numerologies.}
\label{fig:Fig6}
\end{figure*}

Interference estimations are done for each of the used subcarriers separately. Monte Carlo method is applied to increase the statistics in simulation results. The number of tests is 500 and different set of random data is used in each of these tests. Thereafter, the average interference on the subcarriers are estimated. There are four cases in the simulation results presented in \prettyref{fig:Fig6}. Number of usable subcarriers are half of the \ac{ifft} sizes in each case.

In \prettyref{fig:Fig6}, \ac{ini} results are plotted like that there is not any guard bands between the edge subcarriers of two numerologies when there are actually guard bands. The reason of this representation is to make a comparison with different amount of guard bands easily. The below basic inferences are made from the simulation results:
\begin{enumerate}
  \item There is more \ac{ini} at the edge subcarriers of different numerologies.
  \item \ac{ini} present at each subcarrier decreases as the guard band between different numerologies increases.
  \item The effect of guard bands are more prominent for the edge subcarriers.
  \item CP addition causes additional interference for the numerology with smaller \ac{scs}.
  \item Subblocks of the second numerology are constituted by dividing the composite signal. Hence, the symbols of the first numerology causes an extra interference on the second numerology at the receiver side.
  \item \ac{ini} on every $(2^k)$th subcarrier is less than that of the other subcarriers for the numerology with smaller \ac{scs}.
\end{enumerate}

Simulation results show that there are opportunities for the adaptive algorithm designs in 5G as mentioned in \prettyref{sec:opportunity}.

\section{Conclusion}
\label{sec:conclusion}

Next generation communications systems including \ac{nr} are evolving towards increased flexibility in different aspects. Enhanced flexibility is the key to address diverse requirements. Spectral guards and pulse shapes are critical part of this flexibility. These are left for the implementation as long as it is transparent to the counterpart of the communications. \ac{nr} flexibility can be exploited by finding optimal and practical solutions for implementation dependent parts of the 5G standardization. The flexibility of \ac{nr} brings too many open-ended research opportunities compared to the previous cellular communications generations.


\end{document}